\documentclass{aa}  
\usepackage{graphicx}
\usepackage{amssymb}
\usepackage{txfonts}
\usepackage{mathrsfs}
\begin{document}
   \title{Li-rich RGB stars in the Galactic Bulge}
   
   \author{O. A. Gonzalez$^{1,2}$  \and   M.    Zoccali$^{1}$   \and
   L. Monaco$^{3}$  \and   V. Hill$^{4}$ \and S. Cassisi$^{5}$ \and D. Minniti$^{1,6}$  \and
   A. Renzini$^{7}$  \and B. Barbuy$^{8}$ \and  S. Ortolani$^{9}$  \and
   A. Gomez$^{10}$}
   
   \offprints{O. A. Gonzalez}
   \institute{$^{1}$Departamento    Astronom\'ia    y   Astrof\'isica,
   Pontificia Universidad  Cat\'olica de Chile,  Av. Vicu\~na Mackenna
   4860,         Stgo.,         Chile\\         \email{oagonzal@uc.cl;
   mzoccali@astro.puc.cl; Dante@astro.puc.cl}\\ $^{2}$European
   Southern Observatory, Karl-Schwarzschild-Strasse 2, D-85748 Garching, Germany\\
   $^{3}$Departamento   Ciencias  Fisicas  y
   Astronomicas,   Universidad  de  Concepcion,   Concepcion,  Chile\\
   \email{lmonaco@astro-udec.cl}\\
   $^{4}$Universit\'e de Nice Sophia Antipolis, CNRS, Observatoire de la C\^{o}te d'Azur, B.P. 4229, 06304 Nice Cedex 4, France\\ \email{vanessa.hill@obspm.fr}\\
   $^{5}$INAF-Osservatorio Astronomico di Teramo, via M. Maggini, 64100 Teramo, Italy\\ \email{cassisi@oa-teramo.inaf.it}\\
   $^{6}$Specola Vaticana, V00120 Citta’ del Vaticano, Italy\\
   $^{7}$INAF-Osservatorio Astronomico di  Padova, Vicolo dell'Osservatorio  2,I-35122 Padova, Italy\\  \email{alvio.renzini@oapd.inaf.it}\\
   $^{8}$Universidade de   S\~{a}o   Paulo,   IAG,   Rua  do   Mat\~{a}o   1226,   Cidade
   Universit\'aria,     S\~{a}o      Paulo     05508-900,     Brazil\\ \email{Barbuy@astro.iag.usp.br}\\ $^{9}$Universit\`a di Padova,   Dipartimento  di  Astronomia,   Vicolo  dell'Osservatorio 5,I-35122 Padova, Italy\\ \email{sergio.ortolani@unipd.it} \\   
   $^{10}$Observatoire de Paris-Meudon, 92195   Meudon    Cedex,   France\\   \email{Ana.Gomez@obspm.fr}\\ 
}

   \date{Received 05 May 2009; Accepted 09 July 2009}

  
\abstract  
%  context (optional)  
%  aims
{}
{We present  Lithium abundance determination  for a sample of  K giant
stars in the galactic bulge. The  stars presented here are the only 13
stars with detectable  Lithium line (6767.18\AA{}) among $\sim$400 stars for
which  we have  spectra  in this  wavelength  range, half  of them  in
Baade's Window ($b=-4^\circ$) and half in a field at $b=-6^\circ$.}
% methods
{The stars were observed with the GIRAFFE spectrograph of FLAMES@VLT,
with a spectral resolution of R$\sim$20,000. Abundances were derived
via spectral synthesis and the results are compared with those for 
stars with similar parameters, but no detectable Li line.}
% results
{We find 13 stars with a detectable Li line, among which 2 have abundances $\rm A(Li)>2.7$. No clear correlations were found between the Li
abundance and those of other elements. With the exception of the two most Li 
rich stars, the others follow a fairly tight $\rm A(Li)-T_{\rm eff}$ correlation.}
% conclusion
{It would seems that there must be a Li production phase during the
red giant branch (RGB), acting either on a very short timescale, or 
selectively only in some stars. The proposed Li production phase
associated with the RGB bump cannot be excluded, although our targets
are significantly brighter than the predicted RGB bump magnitude for
a population at 8 kpc.}
             
\authorrunning{O.A. Gonzalez et al}
\titlerunning{Li-rich RGB stars in the Galactic Bulge}

\maketitle

%
%________________________________________________________________

\section{Introduction}

The cosmic evolution of Lithium has been a matter of debate in the
recent past, due to some marked inconsistencies between its predicted
abundance and a few key observations (see, e.g., Cyburt et al. 2008 for a 
recent review).

The dominant lithium isotope is $^7$Li,which is thought to be produced
by Big Bang nucleosynthesis, up to abundance A(Li)$\sim$2.2, i.e., the
so-called  Spite   plateau  abundance  observed  in   the  surface  of
metal-poor  stars.  Since  then, Li  has been  produced by  hot bottom
burning  (Sackmann \&  Boothroyd,  1999) during  the asymptotic  giant
branch (AGB) evolution of intemediate mass stars ($>4 \rm M_{\odot}$),
or by cosmic  ray spallation, but also destroyed  in stellar interiors
where tempratures are in excess of $2.0\times 10^{6}$K.

In this context, a typical  bulge star of roughly solar metallicity is
expected to begin its life, on  the main sequence, with a Li abundance
close  to  A(Li)=3.0.  During  the  first  dredge  up,  when  envelope
convection penetrates  down to high temperature regions,  Li is partly
destroyed (diluted)  by $\sim$1.5 dex  (Iben 1967a,b) so  that typical
near   solar   metallicity   RGB    stars   are   expected   to   have
A(Li)$<$1.5. However,  stars on the upper RGB  are expected to
be more Li  depleted due to the extra-mixing  observed to start acting
at the RGB bump (Gratton et  al. 2000, Lind et al. 2009b). In marked
contrast with this  expectation, several Li rich red  giants have been
found to date, both in clusters and in the field (e.g., McKellar 1940;
Faraggiana 1991;  Smith et al. 1995;  Hill \& Pasquini  1999; Kraft \&
Shetrone 2000; Dom\'\i nguez et al. 2004; Monaco \& Bonifacio 2008).

As a  possible explanation  for this evidence,  a Li  production phase
during the  RGB via Cameron-Fowler mechanism (Cameron  \& Fowler 1971)
has  been   proposed  (Sackmann  \&  Boothroyd   1999,  Charbonnel  \&
Balachandran 2000, Denissenkov \& Herwig 2004). In order to synthetize
Li  in   stellar  interiors,  two  conditions   are  required.  First,
temperature    should    be    hot    enough    for    the    reaction
$^{3}$He($\alpha,\gamma$)$^7$Be  to occur.  Second,  $^{7}$Be must  be
quickly transported to cooler regions  where Li can be produced by the
$^{7}$Be($\rm  e^{-},\nu$)$^{7}$Li  reaction.   Thus,   the  reaction
producing Be  should occur very  close to some convective  region, or,
alternatively,   convection  should   penetrate   into  some   burning
shell.  Then,  in low  mass  stars,  some  kind of  extra-mixing  must
circulate the material from the  convective envelope to a region close
to the  H-burning shell. The  source of this  extra-mixing has
been  related  to  shear  inestabilities,  meridional  circulation  or
diffusion.

Charbonnel \& Balachandran (2000) show that a significant number of Li
rich  giants appear  to be  located  close to  the RGB  bump where  an
extra-mixing process is observed  to strongly affect the abundances of
C, $\rm^{12}C/^{13}C$ and  N (Gratton et al. 2000,  Gratton, Sneden \&
Carretta 2004). Based on this evidence, they propose a scenario where,
while the  H-burning shell  erases the molecular  weight discontinuity
left by  the penetration of  the convective envelope during  the first
dredge up, the  extra-mixing could circulate the material  in order to
produce Li. Once  mixing proceeds long enough to  produce the observed
dip in  the carbon isotopic ratio  during and after the  RGB bump, the
fresh $\rm  ^7Li$ would  be quickly destroyed,  such that the  Li rich
phase would be  a very short one. This latter  point would explain why
only  a small  fraction  of the  observed  RGB stars  are actually  Li
rich. However, so far this qualitative  explanation could not
be  reproduced   by  models  accounting   for  rotational  distortions
(Palacios et al. 2006) and  it is also challenged by some observation
of Li rich  stars significantly brighter than the  RGB bump (Monaco \& Bonifacio 2008, Kraft et al. 1999)

Sackmann \& Boothroyd (1999) suggested that, under certain conditions,
an  increase  in  the  Li  abundance  could  be  produced  by  a  deep
circulation   mechanism  after   the  molecular   weight   barrier  is
erased. Therefore, depending on  the extra-mixing details, Li-rich red
giants may be  found at any location of the  RGB. However, the problem
arises when  comparing the behavior  of other elements,  in particular
$\rm  ^{12}C/^{13}C$, with  the  expected values  from their  proposed
model. In particular, in  their scenario the lower $\rm ^{12}C/^{13}C$
value is only  reached at the tip  of the RGB and not  right after the
RGB bump as observed (Gratton, Sneden \& Carretta 2004).

Alternatively,  Denissenkov \&  Herwig (2004)  proposed  a distinction
between  the extra-mixing at  the RGB  bump and  a so  called enhanced
extra-mixing  induced  by rotation  that  could  be  triggered by  the
spinning up by an external source of angular momentum which then could
increase the Li abundance.

Other explanations for Li rich  red giants have been proposed, such as
some mechanism that might prevent  Li dilution during the first dredge
up,  or contamination  by possible  planets.  However,  both phenomena
would  imply enrichment  of $\rm^9Be$  (in addition  to  $\rm^6Li$ and
$\rm^{11}B$) which is  not usually observed in Li  rich stars (Melo et
al.  2006). Therefore,  the internal Li production during  the RGB via
Cameron-Fowler  mechanism seems  to  be the  most  likely scenario  to
explain low mass Li-rich giants.

Recently,  two different  sources  of extra  mixing have  been
proposed  in  order  to  reproduce  the observed  changes  in  surface
abundances.  Thermohaline  mixing  triggered  by  a  double  diffusive
instability (Charbonnel \& Zahn 2007a) and magneto-thermohaline mixing
(Denissenkov et al. 2009) induced  by magnetic buoyancy. While the way
in which thermohaline  mixing may account for observed  Li rich giants
has not been  yet explained , Guandalini et al.  (2009) has shown that
mixing induced  by magnetic buoyancy might explain  the observation of
both  Li-rich and  Li-poor  stars  along the  RGB.  However, a  deeper
understanding  of magnetic  fields in  low mass  giants as  well  as a
larger  sample of  Li rich  RGB stars  are required  to  confirm these
results.

Here  we  present Li  abundances  for  13  stars with  detectable  Li,
observed in  the context of  our survey of  RGB stars in  the Galactic
bulge  (see  Zoccali et  al.  2008  and Lecureur  et  al.  2007 for  a
description of the whole project).

%__________________________________________________________________
\section{RGB Sample}

\begin{figure}
\begin{center}
\includegraphics[scale=0.36,angle=90]{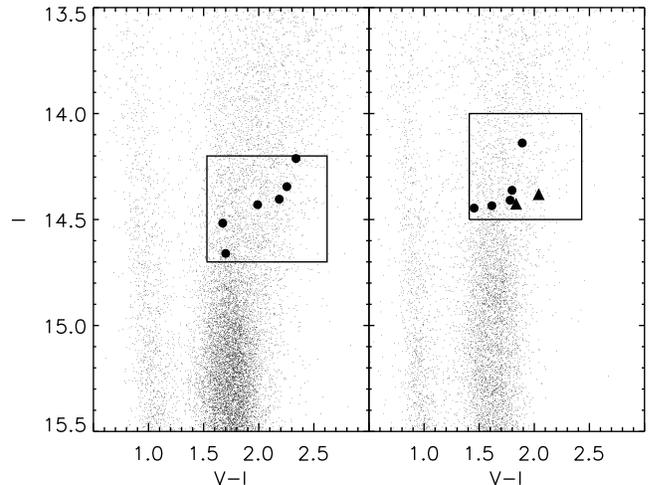}
\caption{Color  magnitude  diagram  for  both fields,  baade's  window
(\textit{left}) and b=-6 (\textit{right}) with the positions of the Li
rich stars  from our  sample marked as  large filled circles.  The two
stars  showing the  higher Li  abundances are  marked as  large filled
triangles. Magnitudes  were obtained  from OGLE catalogue  (Udalski et
al.   2002)  and Zoccali  et  al.  2008  catalogue obtained  from  WFI
images}\label{cmdfull}
\end{center}
\end{figure}

The  spectra discussed here  have been  obtained in  the context  of a
larger  FLAMES-GIRAFFE  survey  of  bulge  K giants,  in  four  fields
(c.f. Zoccali  et al.  2008). Only  two of the  fields, namely Baade's
Window and the field at  $b=-6^\circ$ have been observed with the HR15
setup, including the Li line at  6707.18 $\AA$. The S/N ranges from 40
to 90 and the resolution  is R$\sim$20,000. In total, we have
measured Li in $204+213=417$ bulge giants, whose location in the color
magnitude diagram  (CMD) is shown in Fig.~\ref{cmdfull}.   Only 13 of
those stars showed  a detectable Li line, thus  the following analysis
will   concentrate  on   them,   that  we   will   call  ``Li   rich''
stars\footnote{Note that the name ``Li  rich'' has been often used for
giants  having  A(Li)$>$1.5.  Only  6  of  our  stars show  such  high
abundances,  but we  extend  here the  name  to all  the stars  having
detectable Li, in  contrast with the other $\sim$400  giants for which
we could see no line at all.}.

\begin{table}
\begin{center}
\caption{Bulge fields characteristics}\label{table:i1}
\begin{tabular}{c c c c c c}
\\[3pt]
\hline
N & Name & \textit{l} & \textit{b} & $R_{GC}$ & E(B-V)\\ 
\hline
1 & Baade's Window & 1.14 & -4.18 & 604 & 0.55 \\
2 & $b=-6^{o}$ & 0.21 & -6.02 & 850 & 0.48 \\
\hline
\end{tabular}
\end{center}
\end{table}

\begin{figure}
\begin{center}
\includegraphics[scale=0.38,angle=90]{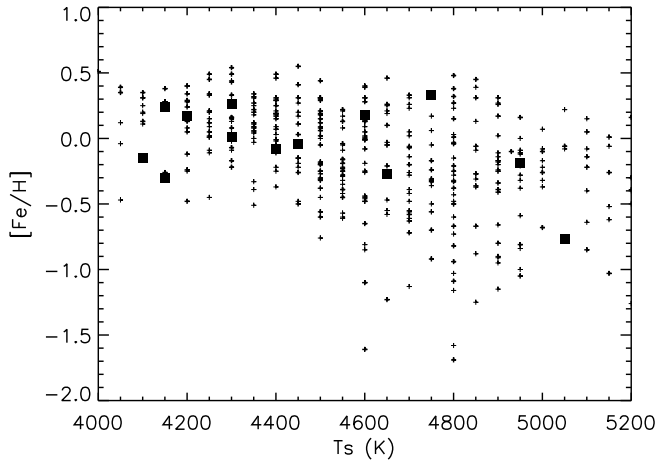}
\caption{Metallicity versus effective temperature diagram for all
the observed bulge giants (small crosses) in both fields.
Li rich giants are shown as big filled squares)
}\label{tempmet}
\end{center}
\end{figure}

Stellar parameters were  obtained from Zoccali et al,  (2008).  As can
be  seen  from   Fig.~\ref{tempmet},  observed  stars  have  effective
temperature  varying from  4000 to  5200 K  and iron  content, [Fe/H],
between $-1.7$  and $+0.5$. Li  rich stars span the  whole temperature
range,  while their  rather high  metallicity is  consistent  with the
small sampling  of a metallicity  distribution peaked close to solar
metallicity.

%_______________________________________________________________________ 
\section{Spectral synthesis and Li abundances}

For  the  13 stars  showing  a  detectable  line, Li  abundances  were
determined  by comparison  with  synthetic spectra  created with  MOOG
(Sneden, 2002).  MOOG is  a FORTRAN code  that performs  spectral line
analysis and spectrum synthesis task under local thermical equilibrium
approximation. MARCS model  atmospheres (Gustafsson 2008) were created
using the stellar parameters given in Table~2.

\subsection{Linelists}
The  TiO  molecular linelist (Plez, 1998) used  by  Lecureur  et al.  (2007)  was
included, but those lines turned out to be  negligible in the relevant
wavelength  range.   CN  lines  are  stronger,   especially  for  cold
stars. The Kurucz CN linelist was obtained from the Kurucz database. Atomic lines in the vicinity of the Li line were obtained from Reddy et al. 2002.

\begin{figure}
\begin{center}
\includegraphics[scale=0.37,angle=90]{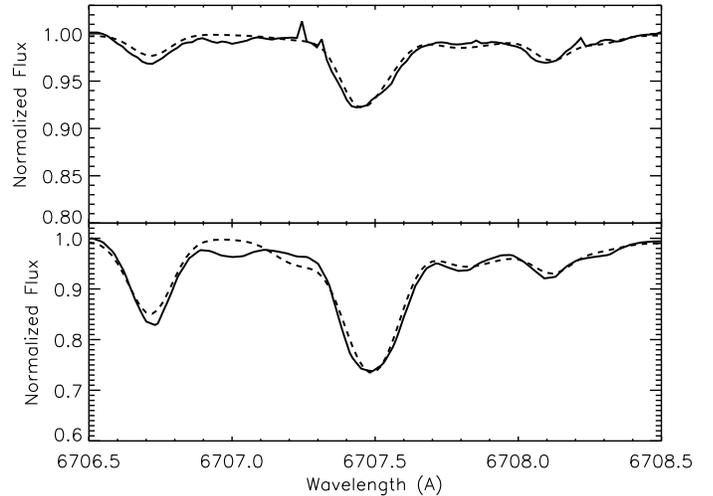}
\caption{Observed spectra as solid line and synthetic spectra as dashed line for Arcturus (upper panel) and for  $\mu$ Leonis (lower panel) used for calibration of the  log  gf values for lines in the vecinity of the Li line}
\label{mol_li}
\end{center}
\end{figure}

Once all linelists were compiled, the log gf values of the CN and atomic lines
within 8 \AA{} from  the  Li line  were  modified  to reproduce  the
observed  spectra   of  Arcturus  and   $\mu$  Leonis,  as   shown  in
Fig.~\ref{mol_li}  using  the  abundances  given  in  Lecureur  et  al
(2007). The log gf value for Ti I line at 6708.025\AA{} was modified from the adopted by Reddy et al. 2002 as well as both V I lines at 6708.125\AA{} and 6708.280\AA{}. The final, calibrated atomic lines used for the synthesis are
listed in Table ~\ref{atomic}.

\begin{table}
\begin{center}
\caption{Atomic linelist in the vicinity of the lithium line, used for this study}\label{atomic}
\begin{tabular}{c c c c}
\\[3pt]
\hline
$\lambda$ ($\AA{}$) & Element & Xiex & log gf\\
\hline  
6707.4331 & Fe I  & 4.610 & -2.300\\  
6707.4500 & Sm II & 0.930 & -3.140\\  
6707.5630 & V I   & 2.740 & -1.530\\  
6707.6440 & Cr I  & 4.210 & -2.140\\  
6707.7400 & Ce II & 0.500 & -3.810\\  
6707.7520 & Ti I  & 4.050 & -2.654\\  
6707.7561 & $^{7}$Li  & 0.000 & -0.428\\
6707.7682 & $^{7}$Li  & 0.000 & -0.206\\
6707.7710 & Ca I  & 5.800 & -4.015\\  
6707.9066 & $^{7}$Li  & 0.000 & -1.509\\ 
6707.9080 & $^{7}$Li  & 0.000 & -0.807\\
6707.9187 & $^{7}$Li  & 0.000 & -0.807\\
6707.9200 & $^{7}$Li  & 0.000 & -0.807\\
6708.0250 & Ti I  & 1.880 & -4.252\\  
6708.1001 & V I   & 1.220 & -3.200\\  
6708.1250 & Ti I  & 1.880 & -2.886\\  
6708.2800 & V I   & 1.220 & -3.178\\  
\hline
\end{tabular}
\end{center}
\end{table}

\subsection{Synthesis}
Synthetic  spectra were  computed  for  each Li  rich  star using  the
corresponding  stellar parameters  and  iron content  listed in  Table
\ref{stellar}.   The  abundances  of  C,N  and O  were  obtained  from
Lecureur et al. (2007), where,  based on a subsample of stars observed
at  higher resolution  with UVES,  approximate relations  were derived
between the abundances of CNO  and [Fe/H].  All other abundance ratios
were scaled to the stellar  metallicity assuming a solar mix (Grevesse
\& Sauval, 1998).   The observed spectra were corrected  by the radial
velocities obtained from DAOSPEC (Stetson \& Pancino 2008) and checked
for  telluric lines  in  the  Li range.  Each  synthetic spectrum  was
compared to the  observed one in two steps.  First,  a 10 \AA{} window
was used  to perform a  correct normalization to the  continuum around
the Li line. Second, a smaller  window (4 \AA{}) was used to reproduce
the Li  line, iteratively  modifying only the  Li abundance  until the
best fitting value was reached. Additionally, NLTE corrections
were calculated by  interpolation the in grids by  Lind et al. (2009),
for   the  stellar   parameters   of   each  one   of   the  Li   rich
stars\footnote{NLTE corrections in Lind et al. 2009 are only available
up  to  solar   metallicity.  Therefore,  corrections  for  supersolar
metallicity stars where calculated assuming [Fe/H]$=$0}.

\begin{figure}
\begin{center}
\includegraphics[scale=0.37,angle=90]{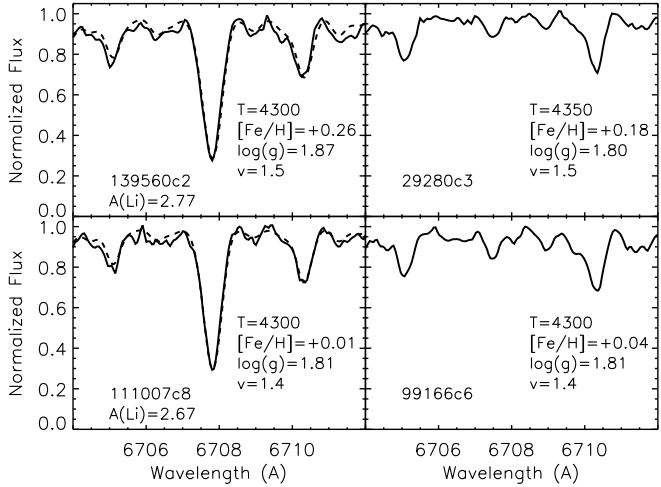}
\caption{\textit{Left  panels:}   Observed  (solid   line)  and
synthetic (dashed line)  spectra for the  two most  Li-rich stars
found from the sample. \textit{Right panels:} Observed spectra for two
stars with similar  $\rm T_{\rm eff}$, $\rm [Fe/H]$, $\rm v_{t}$ and log  g to both Li
rich stars, but no $^{7}$Li line present.}\label{mostli}
\end{center}
\end{figure}

The derived LTE and NLTE Li abundances for the 13 Li rich stars are listed in the
lasts two columns of Table \ref{stellar}.  The two stars showing extremely
high Li abundances, A(Li)$\sim$2.7, are showed in  the left panels of Fig.~\ref{mostli}
along with the best fitting synthetic spectrum. Additionally we performed
spectral synthesis  to other six  stars from each field,  spanning the
whole  parameter range  of  the Li  rich  sample, in  order to  obtain
reference upper limit Li abundance for {\it normal} stars.

The errors on the measured Li abundances were obtained by varying each
of the  stellar parameters by its estimated  error, and re-determining
the  Li  abundance.  The  largest  uncertainty  is  associated to  the
effective  temperature, whose  error  is estimated  to  be $\pm  200$K
(Zoccali et al. 2008),  implying a $\Delta$A(Li)$\sim$0.25 dex for the
hottest of our stars (see Table~\ref{errors}).

\begin{table*}
\begin{center}
\small
\caption{Li Abundance errors associated to the uncertainties in stellar parameters}\label{errors}
\begin{tabular}{c c c c c c c c c}
\\[3pt]
\hline
T & \multicolumn{2}{c}{$\Delta$ T} & \multicolumn{2}{c}{$\Delta$ log $g$} & \multicolumn{2}{c}{$\Delta$ $v_{t}$} & \multicolumn{2}{c}{$\Delta$ [Fe/H]} \\ 
(K) & -200K &+200K& -0.3dex & +0.3dex & -0.2km/s & +0.2km/s & -0.2dex & +0.2dex\\
\hline
4200 & -0.15 & 0.20  & -0.01 & 0.02 & 0.00 & 0.00 &  0.15 & -0.14\\
4650 & -0.20 & 0.20  & -0.01 & 0.01 & 0.00 & 0.00 &  0.20 & -0.16\\
4950 & -0.25 & 0.25  & -0.01 & 0.01 & 0.00 & 0.00 &  0.23 & -0.25\\
\hline
\end{tabular}
\end{center}
\end{table*}

\begin{table}
\begin{center}
\addtolength{\tabcolsep}{-3pt}
\caption{Stellar parameters and determined Li abundance to the sub-sample 
of stars with a distinguishable Li line }\label{stellar}
\begin{tabular}{c c c c c c c c}
\\[3pt]
\hline
N & ID & $\log g$ & vt & Ts & [Fe/H] & A(Li)$_{LTE}$ & A(Li)$_{NLTE}$\\
\hline
1  & 82831 & 1.99 & 1.4  & 4750 & 0.33 & 1.74   &  1.98 \\
2  & 205356 & 2.16 & 1.5 & 4950 & -0.19 & 1.87  &  2.04 \\
3  & 564760 & 1.74 & 1.3 & 4150 & -0.30 & 0.77  &  1.08 \\
4  & 564789 & 1.70 & 1.2 & 4100 & -0.15 & 0.66  &  1.00 \\
5  & 231128 & 1.76 & 1.4 & 4200 &  0.17 & 0.80  &  1.14 \\
6  & 392931 & 1.89 & 1.5 & 4450 & -0.04 & 1.13  &  1.45 \\
7  & 108191c7 & 2.11 & 1.5 & 5050 & -0.77 & 2.17 & 2.24 \\
8  & 69986c2 & 2.01 & 1.5 & 4650 & -0.27 & 1.21  & 1.46 \\
9  & 139560c2 & 1.87 & 1.5 & 4300 &  0.26 & 2.77 & 2.78 \\
10 & 103413c6 & 1.65 & 1.4 & 4150 &  0.24 & 1.01 & 1.34 \\
11 & 75601c7 & 1.79 & 1.5 & 4400 & -0.08 & 0.78  & 1.11 \\
12 & 77419c7 & 1.93 & 1.5 & 4600 &  0.18 & 1.54  & 1.82 \\
13 & 111007c8 & 1.81 & 1.4 & 4300 &  0.01 & 2.67 & 2.68 \\
\hline                                  
\end{tabular}
\end{center}
\end{table}

%_______________________________________________________________________ 
\section{Possible origin of a high Li abundance}

\subsection{Correlations with temperature and metallicity}
In an attempt to  understand the reason for the  high Li observed in
13  of our  stars, we  investigate possible  correlations  between the
derived Li content and the stellar parameters.

From the location of the Li rich stars in the CMD (Fig.~\ref{cmdfull}) we see that they span the full color range of the other sample stars, and they do not clump in magnitude either. This information,
coupled with their uniform spatial distribution -and similar number
in both fields- allows us to exclude that they might belong to a 
different, peculiar, stellar population such as a star cluster.

\begin{figure}
\begin{center}
\includegraphics[scale=0.37,angle=90]{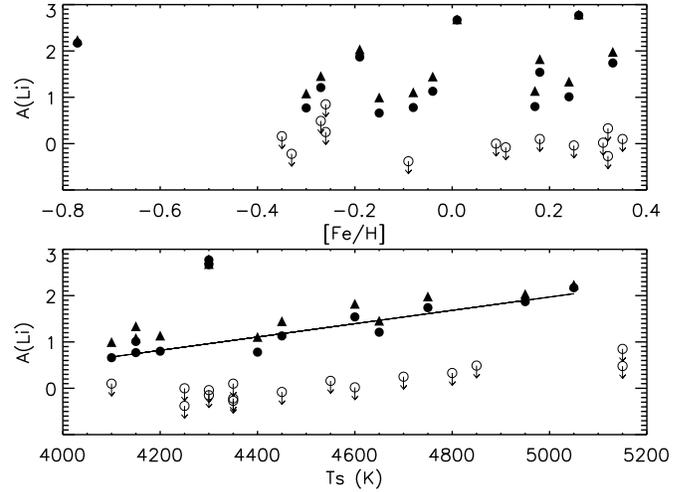}
\caption{Li$_{LTE}$ (filled circles) and Li$_{NLTE}$ (filled triangles) abundances and  stellar  parameters for  the  Li  rich stars. Upper limits in A(Li) for a sample of random stars are shown as
empty  circles. No correlation is seen between lithium  abundances and
metallicity  (upper panel).  A clear  correlation is  visible between
A(Li) and $\rm T_{\rm eff}$ (lower panel) with the only two outliers being
the most  Li rich stars.  A linear fit to the data gives a slope of
0.14$\pm$0.02 dex/100 K.}
\label{parameters}
\end{center}
\end{figure}

All our Li rich stars have metallicities in excess of [Fe/H]$\sim -0.3$
with the exception of one star at [Fe/H]=$-0.77$ (Fig.~\ref{parameters}).
Given the global iron distribution function in these two fields (Zoccali 
et al. 2008) this is entirely consistent with (small) random sampling.

The temperature distribution of Li rich stars is extremely uniform 
across the whole range sampled by our targets, from 4000 to 5200 K.
Very interestingly, with the exception of the two extremely Li rich 
stars shown in Fig.~\ref{mostli}, all the other stars show a rather
tight correlation between A(Li) and $\rm T_{\rm eff}$, as already found
by Brown et al. (1989) in a survey of 644 giants in the DDO 
photometric catalog, Pilachowski et al. (1986) in NGC 7789 and 
Pilachowski et al. (1988) in M67 giants.

In  order to  ensure that  the observed  relation is  real and  is not
produced by the reduced sensitivity to Li detections at hotter temperatures (thereby introducing correlated errors in Teff and A(Li)), we analyse here the effect that an 
error in temperature might introduce on the derived abundances. \\
 
Table \ref{errors} lists the variation of A(Li) due to an error of
200 K in the temperature of the adopted model atmosphere. It appears
that an increase in temperature would imply an increase in the
derived A(Li), and vice versa, thus artificially introducing a positive
slope in the lower panel of Fig.~\ref{parameters}. 
According to Magain (1984), the slope due to correlated errors would
be given by:

\begin{center}
$<\delta f>=c\frac{\sigma^2_{\gamma}}{\sigma^2_{T}}$
\end{center}

where:

$c=\frac{\partial  \rm A(Li)}{\partial \rm T}=0.25$   dex/200K
  describes  how A(Li) is   correlated  with T$_{\rm eff}$, assuming
  the maximum variation, obtained for $\rm T>4950$;

$\sigma_{\gamma}=200$ K is the typical error in effective temperature;

and $\sigma_{T}=333$ K is the variance of the effective temperature range\\

This gives an expected error in  the slope due to correlated errors of
$<\delta f>\sim0.1$ dex/200K. The best fit for the data gives as slope
of  0.28  dex in  200  K.  Therefore most  of  the  observed trend  is
certainly real.

Considering the hypothesis proposed Charbonnel \& Balachandran (2000),
the RGB  bump should  act as  the evolutive instance  in which  we can
separate the stars  in three groups according to  their Li content. i)
Stars under standard dilution should be less evolved than the RGB bump
and therefore could show  the observed trend in Fig. \ref{parameters};
ii) stars  under fresh  Li production which  should be located  at the
first instances  of the RGB bump  before full mixing  takes place; and
iii) stars  which have left the RGB  bump and no Li  content should be
observed on their  surfaces. Thus, Li rich stars  should be located,
in the CMD,  very close to the RGB bump. However,  our stars, being in
principle highly  more evolved than  RGB bump stars, obviously  do not
fit this scenario.
In this context, the  observed decline  in  lithium with  temperature  observed in  our
stars, can only be interpreted as the result of a process occurring as
they evolve through the giant branch.

\subsection{The evolutionary status of the Li-rich stars}
Two of  the stars have A(Li)$\sim$2.7  and they fall  always above the
observed trend. Similar abundances of Li have been previously found in
two  giants by  Brown  et  al.  (1989).   According  to Charbonnel  \&
Balachandran (2000) the latter two stars belong to the RGB bump. Since
they also showed  Be and $^{6}$Li depletion, they  were interpreted as
having recently undergone  a Li production phase.  The  fact that they
had normal  $\rm ^{12}C/^{13}C$ ratios,  then, was thought  to support
the idea  that Li production  should precede the  extra-mixing process
lowering the  carbon isotopic ratio.  Our stars are brighter  than the
expected  location of  the RGB  bump, if  they are  at the  bulge mean
distance. Therefore,  the only  way for them  to fit in  this scenario
would be that they are in fact much closer, at $\sim$ 5 kpc instead of
8 (see below).

Our target  selection box in the  CMD is located  about 0.7 magnitudes
above the horizontal branch red  clump, at $\rm V_{\rm box}\sim 16.1$.
For  a   solar  metallicity  population,  the   expected  $\Delta  \rm
V^{bump}_{HB}$ is  0.5 magnitudes (Zoccali  et al.  1999).   Since the
observed red clump is at $V\sim  16.8$, the RGB bump should then be at
$V\sim17.3$,  i.e.,  1.2  mag  fainter  than  our  target  box.   Even
considering  that the  RGB bump  occurs at  brighter  luminosities for
metal poor stars, not even for the most metal poor Li rich star in our
sample,  the  observed  magnitude  is  compatible  with  the  expected
location of the RGB bump. Assuming  a mean bulge distance of 8 kpc, Li
rich stars in our sample should be  at $\sim 4.6$ kpc from us in order
to actually belong to the RGB bump.  The bulge density at 3.4 kpc from
the galactic center is very low (Rattenbury et al.  2007).  Therefore,
although this possibility cannot be excluded, it is rather unlikely.

\begin{figure}
\begin{center}
\includegraphics[scale=0.35,angle=90]{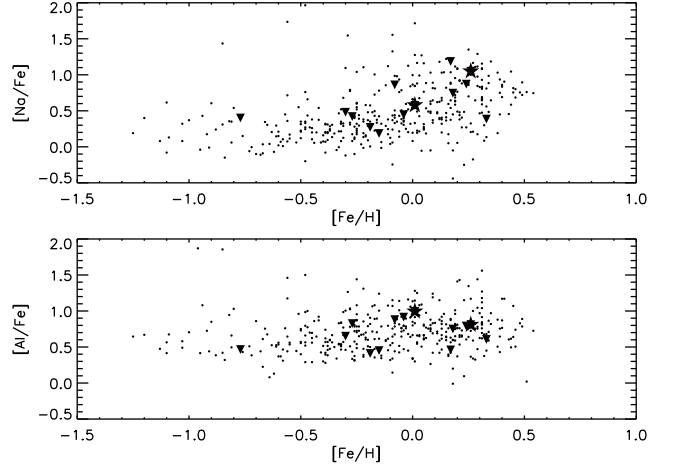}
\caption{[Na/Fe]  and  [Al/Fe] ratios  versus  metallicity. The full  sample is ploted
  as small circles, the Li rich  stars as filled triangles and the two
  most Li rich stars as filled stars.}
\label{alna}
\end{center}
\end{figure}

More likely, these stars might be  {\it disk} RGB bump stars, at $\sim
4.6$ kpc  from us. In  principle one way  to separate bulge  from disk
stars  might  be their  alpha  element  abundances. It has been found that [$\alpha$/Fe] ratios appear to be higher for bulge stars than those belonging to the disk (e.g. Zoccali  et
al. 2006, Lecureur  et al. 2007). If oxygen cannot be measured at the
resolution of GIRAFFE, we can measure Mg, Al, and Na and check if they
are lower than expected for  bulge stars. Figure \ref{alna} shows that
this is not the case, at least for Al and Na.

\subsection{Binary nature}

\begin{figure}
\begin{center}
\includegraphics[scale=0.36,angle=90]{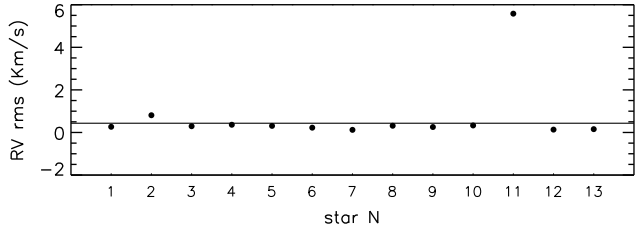}
\caption{Radial velocity variations for the Li rich stars (filled circles) compared to the mean variations of the rest of the sample (solid line)}
\label{radial}
\end{center}
\end{figure}

Costa et al. (2002) found that the Li abundances in binary systems
including a giant show a trend with temperature similar to the one
shown in Fig.~\ref{parameters}. This fact was interpreted as tidal
interactions influencing standard dilution. In fact, binary systems 
with synchronized rotational and orbital motions show higher lithium 
compared to non-synchronized binaries. In order to check for binarity
in our targets, we looked for radial velocity variations among the
individual spectra, before co-addition. For each observed field, 6 to 8 individual spectra were obtained for different sub-samples, spanning a time window from 30 to 70 days. Star 108191c7 is the only one belonging to a sub-sample observed in only two nights. Figure \ref{radial} shows  
the radial velocity  RMS for the Li rich stars, compared to stars 
with no Li line  detected (solid line). Only star 75601c7 shows variations of $\sim$ 
6 km/s, about 20 times higher than the others. Due to small sampling
and projection effects, we cannot exclude that some other stars are
binaries for which we do not detect the variations. However, once
again, binarity for {\it all} the 13 stars would not seem to be the
favoured scenario, from the present data.

\subsection{Circumstellar material and Infrared excess}

In the previous sections we have demonstrated that the Li rich stars
we detect are most likely above the RGB bump. In agreement with that,
also Kraft et al. (1999) detected an enhanced Li star in M3 (star IV-101
with A(Li)=4.0) located 2.25 magnitude brighter than the RGB bump. 
Monaco \& Bonifacio (2008) also found two stars with A(Li)$>$3.5-4.0 
close to the RGB tip of the Sagittarius dwarf spheroidal tidal stream.

De la Reza  et al. (1996,1997) proposed a different scenario for the Li  production along the RGB, associated to
short  episodes of mass  loss occurring  just after  each extra-mixing
event.  In this  way, some fresh Li would  be produced when convection
penetrates close to the H-burning  shell, which might occur at the RGB
bump or brighter.  The thin layer containing fresh Li would be quickly
carried to the surface (and observed) but it would be lost by the star
right afterwards.  Observational support to this scenario was given by
the detection of infrared excess in some Li-rich giant stars (Gregorio
-Hetem et al.  1993, De la Reza et al 1996,1997, Castilho et al. 1998).

Later on, however, Jasniewicz  et  al. (1999) analyzed 29 giants with 
infrared excess and did not find any correlation with lithium abundance.
In particular, most of their target stars show no detectable Li, 7 of
them have abundances compatible with standard dilution (A(Li)$\sim$1.5),
and only one has A(Li)=3.0, the latter being indeed close to the RGB
bump.

\begin{figure}[]
\begin{center}
\includegraphics[scale=0.37,angle=90]{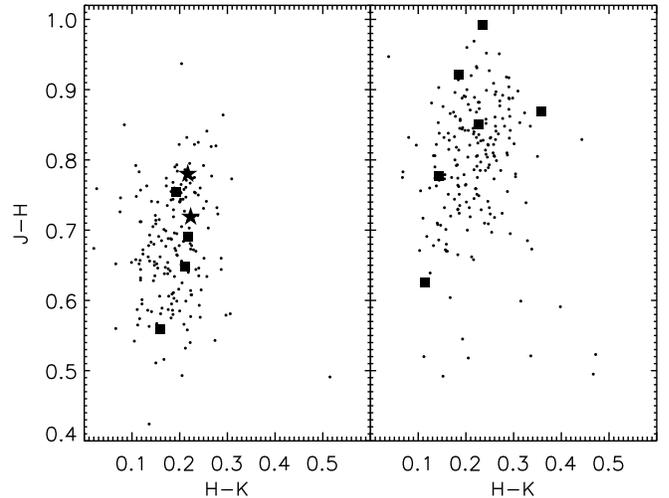}
\caption{(J-H,H-K) Colour-colour diagram for the  complete  sample of  stars in  the
$b=-6^\circ$ field (left) and  Baade's Window (right) including the Li
rich stars (filled  squares). The two most Li  enhanced stars 139560c2 and
111007c8 are plotted as filled stars.
}\label{excess}
\end{center}
\end{figure}

Ideally,  mid or  far  infrared  data are  needed  to detect  infrared
excess. Since at the moment we do not have such data, we have examined
the (J-H,H-K) colour-colour diagram to look for an excess $K$ brightness. As shown
by Jones (2008),  stars showing a clear excess  flux ($\gtrsim$ 1 mag)
at 8$\mu m$ with respect to  $K$, also show a $\sim$0.4 mag excess
brightness  in  $H-K$  with  respect to  no-IR-excess  stars.   Figure
~\ref{excess} shows that  no near-IR excess is visible  for any of the
Li  rich stars  when  compared to  the  bulk of  normal  stars in  our
sample. However, it  should be noted that this does  not mean that the
stars do not have circumstellar  material. Indeed, as shown Origlia et
al.  (2007), moderate  ($\sim$0.5 mag)  excess  in $K-8\mu m$ is  not
detectable in near IR.

Additionally, a significant number of Li rich giants have been
found to be also rapid rotators (Drake et al, 2002). However, there is
not  a one-to-one  relation  between Li  enrichment  and rotation  (de
Medeiros et al.  1996). Furthermore, as found by  Drake et al. (2002),
this  connection only  seems  to  be present  when  high IR-excess  is
present. Indication of fast rotation ($v$ sin $i$ $\gtrsim$ 8 km/s) is
observed as unusually  broad lines in spectral features.  In our case,
measured FHWM  of Li-rich stars  show no enhancement when  compared to
the average of the sample.

\begin{figure}[]
\begin{center}
\includegraphics[scale=0.37,angle=90]{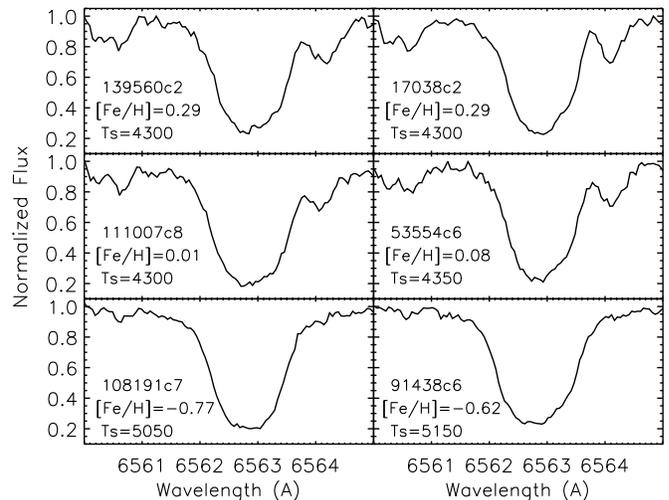}
\caption{$\rm H_{\alpha}$ profile for 3 Li-rich \textit{(left panels)} and
3  Li-poor  \textit{(right  panel)}  stars.  There  are  no  signs  of
chromospheric  activity  in the  profiles,  nor  any other  difference
between Li-rich and Li-poor stars}\label{halpha}
\end{center}
\end{figure}

Finally, were the high  Li abundance related  to circumstellar
material, we could expect asymmetric wings and/or emission component
on the H$_\alpha$ line. Figure~\ref{halpha} shows that this is not
observed, as the H$_\alpha$ of Li rich stars (right panels) is identical to that of
any other star of the sample (left panels).

%_______________________________________________________________________ 
\section{Conclusions}

We have analyzed the Li abundance from a sample of $\sim 400$ K giants
in the galactic  bulge. A sub sample of 13  stars present a detectable
Li line for  which we have measured A(Li)=0.7-2.8. 

The sample stars could be divided in three categories:

i) {\bf $\sim$ 400 normal stars} showing no Li line.

ii) {\bf  11 Li  rich  stars}  with A(Li)=0.66-1.87.  These  have  abundances
compatible with  the standard Li  dilution ocurring at the  1st dredge
up. However they seem to  have somehow avoided the second extra-mixing
episode,  further diluting Li  after the  RGB bump  (see Fig.8  in the
review by Gratton et al. 2004). These stars show a clear
correlation between A(Li) and T$_{\rm eff}$.

iii) {\bf 2 highly Li rich stars} with A(Li)$\sim$2.8. These stars necessarily
underwent a Li production phase.

Most  of  the proposed  explanations  for  the  presence of  (a  small
fraction of) Li  rich RGB stars involve the  Comeron Fowler mechanism,
associated to a deep mixing  episode, occurring either at the RGB bump
(Charbonnel \& Balachandran 2000) or brighter (Sackmann \& Boothroyd 1999, Denissenkov \& Herwig 2004). 
From the position of our stars on the  CMD we conclude that it is rather unlikely that any of
them belongs to the RGB bump, since they are observed to be $\sim$ 1.2
mag brighter than its expected position, at the mean bulge distance.

No clear indication has been found  for the presence of a companion to
these stars, nor for an  infrared excess or some kind of chromospheric
activity.  By dividing the spectrum of Li rich stars by the one of 
a {\it normal} star with similar parameters we could not identify any
other spectral feature that might be different between the two.

We have presented more evidence for the presence of Li-rich stars in the first ascent RGB, contrary to the predictions of canonical stellar evolution. None of the proposed explanations could be confirmed here, even though some of them could not be firmly discarded either. Clearly, our poor understanding of this evidences is strongly affected by the small number statistics, partly due to the intrinsec rareness of the phenomenon but also to the lack of a dedicated survey. Specifically designed observations would be needed, ideally in simple stellar populations, in order to establish clearly the evolutionary status of Li-rich giants.

%%%%%%%%%%%%%%%%%%%%%%%%%%%%%%%%%%%%%%%%%%%%%%%%%%%%%%%%%%%%%%%%%%%%%%%%%%%%%%%%%% 

\begin{acknowledgements}
We thank Alejandra Recio-Blanco, Angela Bragaglia and Patrick de Laverny for very useful comments and discussions. OG thanks Karin Lind for providing us with the interpolation code and grid for NLTE corrections. MZ and OG aknowledge support by Proyecto Fondecyt Regular $\sharp$1085278. MZ and DM are partly supported by the BASAL CATA and the FONDAP Center for Astrophysics 15010003.
%%%%%%%%%%%%%%%%%%%%%%%%%%%%%%%%%%%%% 
%%%%%%%%%%%%%%%%%%%%%%%%%%%%%%%%%%%%%
\end{acknowledgements}


\begin{thebibliography}{}
\bibitem[bensby]{2004}
 Bensby, T., Feltzing, S., \& Lundstrom, I., 2004, A\&A, 415, 155
\bibitem[boothroyd]{1999} 
 Boothroyd, A.I. \& Sackmann, I.J., 1999, ApJ, 510, 232
\bibitem[brown]{1989}
 Brown, J.A., Sneden, C., Lambert, D.L. \& Dutchover, E. Jr., 1989, ApJS, 71, 293
\bibitem[Cameron]{1971}
 Cameron, A.G.W., \& Fowler, W.A., 1971, ApJ, 164, 111
\bibitem[Castilho]{2000} 
 Castilho, B.V., Gregorio-Hetem, J., Spite, F., Barbuy, B. \& Spite, M., 2000, A\&A, 364, 674
\bibitem[Charbonnel]{2000} 
 Charbonnel, C., \& Balachandran S.C. 2000, A\&AS, 359, 563
\bibitem[Charbonnel]{1998}
 Charbonnel, C., \& do Nacimento . 1998, A\&AS, 359, 563
\bibitem[Charbonnelzahn]{2007} 
 Charbonnel, C., \& Zahn, J.P., 2007a, A\&A, 467, L15
\bibitem[Charbonnelzhanb]{2007}
 Charbonnel, C., \& Zahn, J.P., 2007b, A\&A, 476, L29
\bibitem[delareza]{2000}
 de La Reza, R., da Silva, L., Drake, N. \& Terra, M., 2000, ApJ, 115, 117
\bibitem[delareza]{1995}
 de La Reza, R. \& da Silva, L., 1995, ApJ, 439, 917
\bibitem[demedeiros]{1996}
 de Medeiros, J.R., Melo, C.H.F. \& Mayor, M., 1996, A\&A, 309, 465
\bibitem[Denissenkov]{2004}
 Denissenkov, P.A. \& Herwig, F., 2004, ApJ, 612, 1081
\bibitem[Denissenkov]{2009}
 Denissenkov, P.A., Pinsonneault, M. \& MacGregor K.B., 2009, ApJ, 696, 1823
\bibitem[Dominguez]{2004}
 Dom\'inguez, I., Abia, C., Straniero, O., Cristallo, S., \& Pavlenko, Y.V., 2004, A\&A 422, 1045
\bibitem[Drake]{2002} 
 Drake, N.A., de la Reza, R., da Silva, L., Lambert, D.L., 2002, AJ, 123, 2703
\bibitem[Faraggiana]{1991} 
 Faraggiana, R., Gerbaldi, M. Molaro, P., \&, Lambert, D.L., 1991, MmSAI, 62, 189
\bibitem[Grevesse]{1998}
 Grevesse, N., \& Sauval, A.J., 1998, Space Sci. Rev., 85, 161
\bibitem[Gratton]{1990}
 Gratton, R.G., \& Sneden, C., 1990, A\&A, 234, 366
\bibitem[Gratton]{2000}
 Gratton, R.G., Sneden, C., Carretta, E., Bragaglia, A., 2000, A\&A, 354, 169G
\bibitem[Gratton]{2004}
 Gratton, R., \& Sneden, C., \& Carretta, E. 2004, ARA\&A, 42, 385
\bibitem[Guandalini]{2009}
 Guandalini, R., Palmerini, M., Busso, M. \& Uttenthaler, S., 2009, 
\bibitem[Hill]{1999} 
 Hill, V., \& Pasquini, L.,  1999, A\&A 348, L21
\bibitem[Iben]{1967}
 Iben I.J., 1967, ApJ, 147, 624
\bibitem[Ibenb]{1967} 
 Iben I.J., 1967, ApJ, 147, 650
\bibitem[Jasniewicz]{1999} 
 Jasniewicz, G., Parthasarathy, M., de Laverny, P. \& Thevenin, F., 1999, A\&A, 342, 831
\bibitem[Jones]{2008} 
 Jones, M.H.,2008, MNRAS, 387, 845
\bibitem[Kraft]{2000}
 Kraft, R.P., \&  Shetrone, M.D., 2000, LIACo 35, 177
\bibitem[Kurucz]{1993}
 Kurucz, R.L., 1993, CD-ROMs, http://kurucz.harvard.edu
\bibitem[Lecureur]{2007}
 Lecureur, A., Hill, V., Zoccali, et al. 2007, A\&A, 465, 799
\bibitem[Lind]{2009a}
 Lind, K., Asplund, M., Barklem, P.S., 2009a, A\&A accepted, {\tt arXiv:0906.0899v1 [astro-ph.SR]}
\bibitem[Lind]{2009b}
 Lind, K., Primas, F., Charbonnel, C., Grundahl, F., Asplund, M., 2009b, A\&A accepted, {\tt arXiv:0906.2876v3 [astro-ph.SR]}
\bibitem[Magain]{1984} 
 Magain, P., 1984, A\&A, 134, 189 
\bibitem[Mckellar]{1940}
 McKellar, A. 1940, PASP, 52, 407
\bibitem[Minniti]{1998}
 Minniti D., Cook K.H., Vandehei T., Alcock C., Griest K., 1998, ApJ, 499, L175
\bibitem[Minniti]{2008}
 Minniti, D. \& Zoccali, M., 2008, IAU symposium, 245, 323
\bibitem[Monaco]{2008}
 Monaco, L. \& Bonifacio, P., 2008, MmSAI, 79, 1
\bibitem[Origlia]{2007}
 Origlia, L, Rood, R.T., Fabbri, S., et al., 2007, ApJ, 667L, 85O
\bibitem[Palacios]{2001}
 Palacios, A., Charbonnel, C., \& Forestini, M., 2001, A\&A, 375, L9
\bibitem[palacios]{2006}
 Palacios, A., Charbonnel, C., Talon, S., Siess, L., 2006, A\&A, 453, 261P
\bibitem[Pilachowski]{2000}
 Pilachowski, C.A., Sneden, C. Harmer, D. \& Willmarth, D., 2000, ApJ, 119, 2895
\bibitem[Pilashowski]{1990}
 Pilachowski, C.A., Sneden, C. \& Hudek, D., 1990, ApJ, 99, 1225
\bibitem[Plez]{1998}
 Plez, B., 1998, A\&A, 337, 495
\bibitem[Pompeia]{2002}
 Pompeia, L., Barbuy, B., Grenon, M. \& Castilho, B.V., 2002, ApJ, 570, 820
\bibitem[Ramaty]{2001}
 Ramaty, 2001, SSRv 99, 51
\bibitem[Recio]{2007}
 Recio-Blanco, A., de Laverny, P., 2007, A\&A, 461, L13
\bibitem[Reddy]{2002}
 Reddy, B.E., Lambert, D.L., Laws, C., Gonzalez, G. \& Covey, K., 2002, MNRAS, 335, 1005
\bibitem[Sackmann]{1999}
 Sackmann, I.,  \& Boothroyd, A.I. 1999, ApJ, 510, 217
\bibitem[Smith]{1995}
 Smith, V.V., Plez, B., Lambert, D.L., \& Lubowich, D.A.,  1995, ApJ, 441, 735
\bibitem[Spite]{1982}
 Spite, F. \& Spite. M. ,1982, A\&A, 115, 357
\bibitem[Stetson]{2008}
 Stetson, P.B. \& Pancino E., 2008, PASP, 120, 1332 
\bibitem[Uttenthaler]{2007}
 Uttenthaler, S., Lebzelter, T., Palmerini, et al., 2007, A\&A, 471, L41
\bibitem[Zoccali]{1999}
 Zoccali, M., Cassisi, S., Piotto, G., Bono, G. \& Salaris, M. 1999, ApJ, 518, L49
\bibitem[Zoccali]{2006}
 Zoccali, M., Lecureur, A., Barbuy, B., et al., 2006, A\&A, 457, L1
\bibitem[Zoccali]{2008}
 Zoccali, M., Lecureur, A., Hill, V. et al., 2008, A\&A, 486, 177

\end{thebibliography}
\end{document}